\documentclass{ws-procs9x6-cpt25}
\graphicspath{{three_Figures/}}
\begin{document}
	
	\newcommand{\refeq}[1]{(\ref{#1})}
	\def\etal{{\it et al.}}

	\title{Dalitz Plot Kinematics\\ for a
		Lorentz-Violating Three-Body Decay}
	
	\author{Joshua\ O'Connor}
	
	\address{Department of Physics and Astronomy, University of South Carolina,\\
		Columbia, South Carolina 29208, USA}

	\begin{abstract}
		We outline the analysis of three-particle interactions and decay processes for leading-order modifications 
		in a Lorentz-violating quantum field theory. 
	\end{abstract}

	\bodymatter
	
	\section{Introduction}
	
	Modern particle theories typically emphasize calculations and modifications thereof to the transition matrix. When considering the inclusion of symmetry-breaking effects, however, the kinematic modifications made in formal scattering theory can be close to, or in many cases even more important than, changes to matrix elements.
	In physical theories, the estimated available phase space can produce useful results; for example, in Ref.~\refcite{ref-lehnert2}
	the exact radiation rate for a Cerenkov-type effect
	with modified field equations was found to agree well with the kinematically allowed region estimate.\cite{ref-lehnert2}
	
	We define the SME framework used in our kinematics calculation. Practicalities of three-body decays are given using a
	description of the Lorentz-violating background tensor $c_{\mu\nu}$ in the modified phase space. Then we give an
	overview of the kinematics in terms 
	of Dalitz plots and propose possibilities for extensions of this type of prescription. 
	
	\section{Lorentz-violating field theory in phase space}
	\label{sec-theory}
	Our model is developed in the fermionic sector of the Standard-Model Extension (SME). With the use of the spin-independent coefficients $c_{\mu\nu}$
	(a symmetric two-index tensor background), the results are equivalent to those in a scalar sector, since our procedure is based purely on modified phase space. 
	In Refs.~\refcite{ref-kost1,ref-kost2}, the basis for any general local field theory that involves Lorentz-violating modifications to physical processes containing known Standard-Model fields
	is derived. The $c_{\mu\nu}$ coefficients describe violations of rotation and boost symmetries, but without any breaking of CPT.
	
	\begin{figure}
		\centering
		\includegraphics[angle=0,width=7cm]{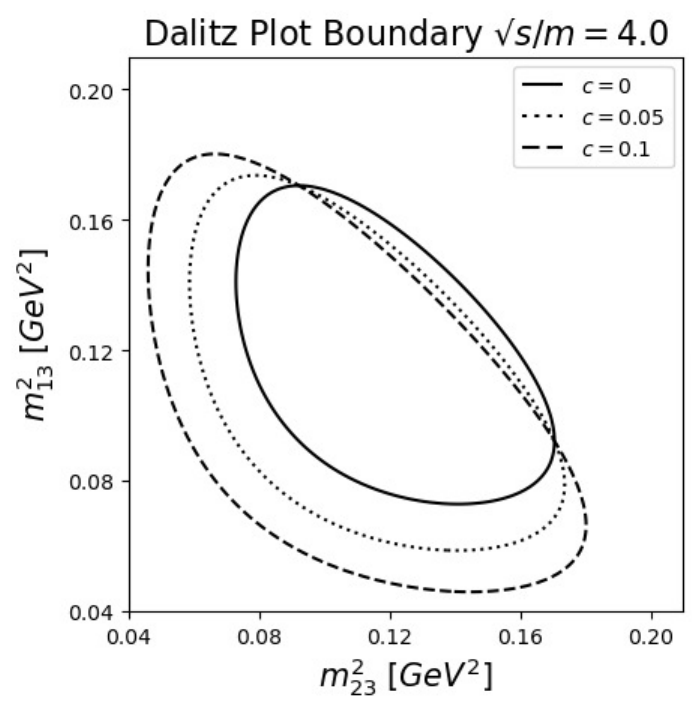}
		\caption{Shape of the Dalitz plot for the $\sqrt{s}/m$ value corresponding to
			$\eta\rightarrow3\pi^{0}$. The standard outline is shown, along with outlines for two nonzero values
			of the Lorentz-violation parameter $c_{\mu\nu}$.}
		\label{fig-Dalitz1}
	\end{figure}
	
	\begin{figure}
		\centering
		\includegraphics[angle=0,width=7cm]{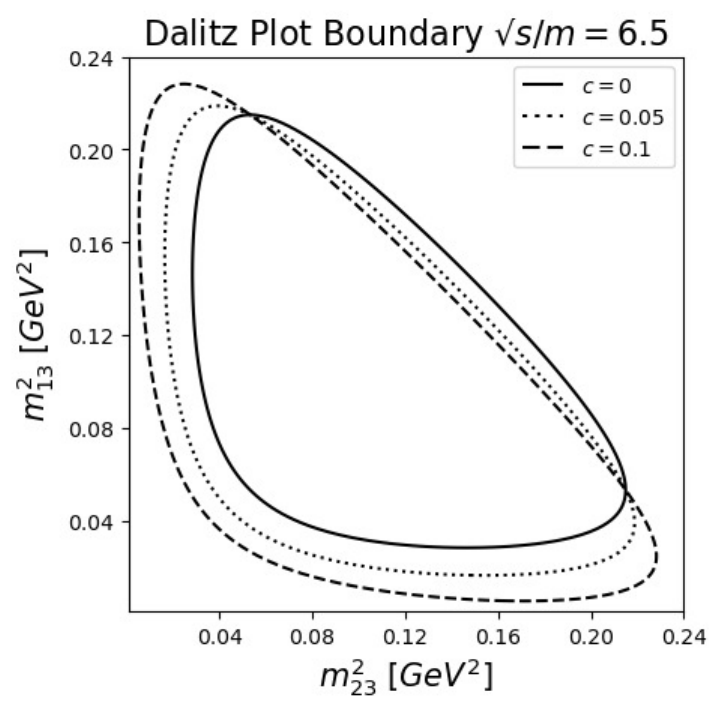}
		\caption{Shape of the Dalitz plot for a $\sqrt{s}/m=6.5$ value.}
		\label{fig-Dalitz2}
	\end{figure}
	
	Consider an $\eta$ meson or kaon as the Lorentz-invariant initial particle, which then decays in the center-of-mass
	frame into three Lorentz-violating spinless particles of equal mass, such as pions.
	A compact description of the kinematics is contained in the dispersion relation given for the pion field:
	\begin{equation}
		\left(g^{\mu\nu}+2c^{\mu\nu}\right)p_{\mu}p_{\nu}-m^{2}=0.
		\label{eq-disp}
	\end{equation}
	The Lorentz-noninvariant modification to the kinematics allows us to calculate the three-particle decay rate with modified particle momenta and energies.\cite{ref-altschul4}
	The main calculation of the decay rate is modified, as opposed to the original Lorentz-symmetric case, by the inclusion in the integrand of modified energy-momentum-conserving Dirac $\delta$-functions. 
	These dispersion relations, kinematics, and further details are laid out in Ref.~\refcite{re-alt+oco2}, the paper
	summarized in this contribution. 
	The leading-order decay rate that results without neglecting any coefficients under a particular
	texture of $c_{\mu\nu}$ modifications simplifies to
	\begin{equation}
		\label{eq-Gamma-c13-not-0}
		\Gamma=\frac{|\mathcal{M}|^2}{512\pi^2\sqrt{s}}\left[s(1+2c)
		+4cm^{2}+4cm^{2}\log\left(\frac{2mc}{\sqrt{s}-2m}\right)\right].
	\end{equation}
	The last term in the expression is recognized as the Lorentz-noninvariant analog of an ultraviolet divergence, with poles at the correct values of the particle mass. 
	Regularization techniques and asymptotic approximations could be further applied to models of this type.\cite{ref-zeta} The leading-order series expansion results in the above expression for the decay rate,
	which depends not only on first-order $c_{\mu\nu}$ coefficients but also on a larger additional logarithmic enhancement
	term of $\mathcal{O}(c\log c)$.    
	
\begin{figure}
		\centering
		\includegraphics[angle=0,width=7cm]{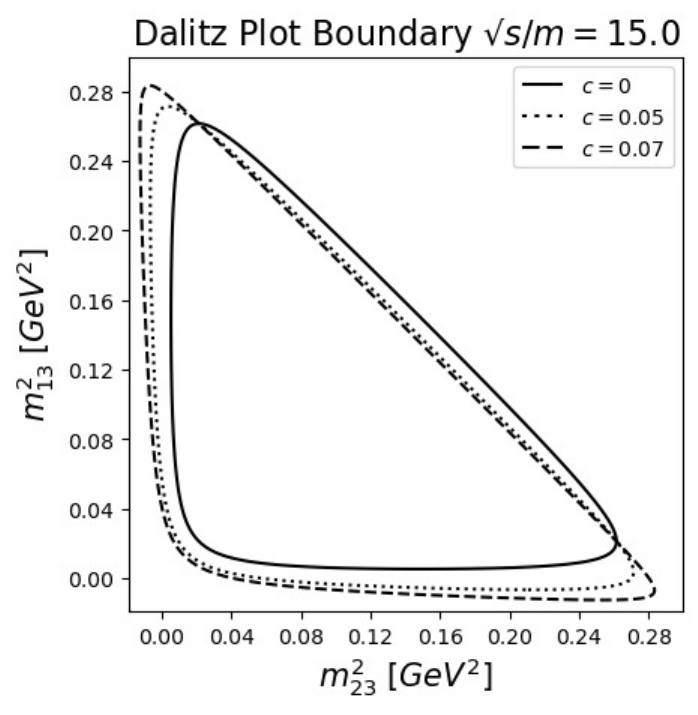}
		\caption{Shape of the Dalitz plot for an ultrarelativistic value of $\sqrt{s}/m=15.0$.}
		\label{fig-Dalitz3}
	\end{figure}

	We can more fully express the kinematics using modified Dalitz-plot boundaries. In this case, the
	elongation of the curve depends only on the available decay energy compared with the common
	mass of the daughters, $\sqrt{s}/m$. If $c_{\mu\nu}$ is nonzero, the plot boundary
	shifts according to the dashed and dotted curves in each plot.
	This is linked to the fact that, under Lorentz transformation, the allowed physical region will be
	rotated and shifted relative to the case without Lorentz violation. Figure~\ref{fig-Dalitz1} shows the effect of $c_{\mu\nu}$ on the shape of the Dalitz plot for a decay
	with $\sqrt{s}/m=4.0$, for the physical decay $\eta\rightarrow3\pi^{0}$.
	Precision measurements of this particular decay
	are already recognized to be important, since the matrix element, calculated using chiral perturbation
	theory, depends directly on the isospin-violating
	up--down mass difference, $m_{d}-m_{u}$.\cite{ref-anisovich,ref-bijnens,ref-kampf1}
	The higher energy and ultrarelativistic plots are shown in 
	Figs.~\ref{fig-Dalitz2} and~\ref{fig-Dalitz3}. 
	Lorentz violation changes the shape of the outline for three-pion decays, with more pronounced fractional changes
	at lower values of $\sqrt{s}/m$.
	
	Increasing the strength of the Lorentz violation (in this specific model),
	shifts the kinematically allowed region toward smaller values of $m_{23}^2$ and $m_{13}^2$---in a symmetric
	fashion, because the daughter particles are identical. 
	In the strongly relativistic regime, as the shape becomes increasingly triangular, we can see that it correspondingly becomes relatively less sensitive to the variety of Lorentz
	violation that we have considered. This appears to be a generic feature of $c_{\mu\nu}$-type Lorentz
	violation. This is perhaps not unexpected, with a $c_{\mu\nu}$ coefficient tending to shift the boundaries
	of kinematically allowed regions by less 
	as the pion masses become less important and the energy-momentum
	relations for the outgoing particles become nearly linear.
	
	\section{Conclusions and Outlook}
	\label{sec-concl}
	
	Calculations in the presence of Lorentz violation increase the number of singularities in the decay integral and 
	boundary equation.
	The anisotropy of Lorentz violation is partially responsible for these complications, since the particle energies now depend 
	on the angular orientations of the daughter particles.
	In our model, whether or not in the presence of nonzero Lorentz-symmetry-breaking coefficients, our plot is
	symmetric, since forming the scalar quantities $m_{13}^{2}$ and
	$m_{23}^{2}$ drops the leading-order directional information. 
	If this type of analysis were applied to weak-sector $\beta$ decays, such as muon decay,\cite{
		ref-vos3,ref-sytema}
	where the daughter particles are not identical scalars, 
	the convex outline could take a more asymmetrical shape. Phase-space modifications of this type could be generalized 
	to a broad class of SME coefficients, such as the charged-fermion sectors.\cite{ref-altschul12,ref-schreck4} 
	
	More elaborate versions of the model used here, with more complicated $c_{\mu\nu}$ textures, could be of use along with time-dependent 
	pion data sets to provide signatures or bounds for Lorentz violation.
	The Dalitz plots can be binned for different sidereal times to find shifts in the physically allowed region as the laboratory rotates with the Earth. 
	There have been some previous considerations
	of how reaction thresholds for particle processes would change with the Earth's rotation, but the Dalitz
	plot provides a more general structure.


	\section*{Acknowledgments}
	The author thanks both B.\ Altschul and M.\ Schindler
	for working with him and for their time in very useful discussions.

\end{document}